\documentclass[prl,superscriptaddress,a4paper,twocolumn,showpacs]{revtex4}
\usepackage{amsmath,amsfonts,graphicx,color}
\usepackage{amssymb}
\def\>{\rangle}
\def\<{\langle}

\def \be{\begin{equation}}
\def \ee{\end{equation}}
\def \beq{\begin{equation}}
\def \eeq{\end{equation}}
\def \bea{\begin{eqnarray}}
\def \eea{\end{eqnarray}}

\def \E{{\cal E}}
\def \F{{\cal F}}

\def \O{{\cal O}}
\def \Tr{\mbox{Tr}}
\begin{document}

\title{Dynamics of a quantum reference frame undergoing selective measurements and coherent interactions}
\author{Mehdi Ahmadi}%
\affiliation{Institute for Mathematical Sciences, Imperial College
London, London SW7 2PG,
United Kingdom}%
\affiliation{Optics Section, Blackett Laboratory, Imperial College
London, London SW7 2BW, United Kingdom}%
\author{David Jennings}%
\affiliation{Institute for Mathematical Sciences, Imperial College
London, London SW7 2PG,
United Kingdom}%
\affiliation{Optics Section, Blackett Laboratory, Imperial College
London, London SW7 2BW, United Kingdom}
\author{Terry Rudolph}%
\affiliation{Institute for Mathematical Sciences, Imperial College
London, London SW7 2PG,
United Kingdom}%
\affiliation{Optics Section, Blackett Laboratory, Imperial College
London, London SW7 2BW, United Kingdom}%

\date{\today}

\begin{abstract}
We consider the dynamics of a quantum directional reference frame
undergoing repeated interactions. We first describe how a precise
sequence of measurement outcomes affects the reference frame,
looking at both the case that the measurement record is averaged
over and the case wherein it is retained. We find, in particular,
that there is interesting dynamics in the latter situation which
cannot be revealed by considering the averaged case. We then
consider in detail how a sequence of rotationally invariant
unitary interactions affects the reference frame, a situation
which leads to quite different dynamics than the case of repeated
measurements. We then consider strategies for correcting reference
frame drift if we are given a set of particles with polarization
opposite to the direction of drift. In particular, we find that by
implementing a suitably chosen unitary interaction after every two
measurements we can eliminate the rotational drift of the
reference frame.

\end{abstract}

\maketitle

\section{Introduction}

It is common to assume the control fields used to manipulate
quantum systems are of infinite strength and therefore classical.
It is possible, however, to relax this assumption, and to treat
them within the quantum formalism \cite{BRSreview} - that is, as
systems of bounded size/strength, and to then investigate the
limitations that this finiteness does or does not impose. From the
perspective of quantum computing this could be desirable because
the inevitable miniaturization of quantum information processing
devices - such as ion trap chips - may make using small strength
control fields a necessity (current proposals would require
hundreds of watts of laser power for a full scale quantum
computation). From the perspective of quantum communication the
issue of finite-sized reference frames raises interesting
questions regarding the fact that the shared references commonly
used by the separated parties can drift, and realigning them
requires further resource expenditure. Finally, there are
interesting foundational reasons for considering finite-sized
references \cite{CHRB1},\cite{BRST2009},\cite{CHRB2}. An example
is the work on finite precision measurements, black hole entropy
and symmetry
deformations\cite{GPP1},\cite{girelli-poulin07}.
Another example, more pertinent to the work to be presented here,
is the work on ``quantum clocks'' - for instance the Page-Wootters
model of a clock which has developed into the the conditional
probability interpretation of time in quantum gravity \cite{PW}.

In this paper we continue a line of investigation
\cite{BRST2007},\cite{BRS2} into a simple model of degradation of
a quantum reference frame consisting a large spin system as it
repeatedly interacts with a series of incoming ``source''
particles. In \cite{BRST} this program of investigation was
initiated by considering a source of unpolarized spin-1/2
particles, each of which has its component of spin measured
against a reference spin directional frame, by implementing the
optimal measurement \cite{BRS2} for determining the relative
direction between the frame and the system. An example of such a
procedure might be the measurement of qubits in a BB84
key-distribution protocol by a finite strength magnetic field. The
conclusion there was that in such circumstances the reference
would be useful for a time (number of uses) that scales
quadratically in the size (ie spin) of the reference. This
conclusion was shown to be quite generally true for rotationally
invariant source particles in \cite{JC}. In \cite{PY} the
investigation was simplified and extended to the case where the
source of particles has some net polarization - such as in a B92
type key distribution for example. An interesting result of
\cite{PY} was that in this instance the drift of the reference
frame was more important to its degradation that the ``diffusion''
caused by the entanglement with the particles, and now the
reference would only be useful for a time linear in its size.

Both \cite{PY} and \cite{BRST} considered the case of
\emph{measuring} the source particles against the reference frame.
The results of \cite{JC} also apply, however, to the case where we
use the reference as a mechanism for doing coherent (unitary)
interactions between the reference and an unpolarized stream of
source particles. In this article we consider the case of
degradation when we do coherent interactions between the reference
system  and a polarized source of particles. We also consider how
well one might correct for the reference frame drift in a simple
model wherein we are given, in addition to the polarized set of
source particles, a smaller number of particles which are known to
have a polarization in a direction opposite to those of the
source. We begin, however, by revisiting the case of using the
reference to implement measurements on a polarized source of
particles, exploring in more detail the dynamics in the case that
the measurement results are \emph{not} averaged over.

\section{Evolution of the reference frame under measurement interactions}

We briefly introduce the formalism for our investigations by
recapping the case of a directional quantum reference frame (QRF)
used for measurement; in the main we are following the formulation
of \cite{PY}.

In standard quantum measurement schemes, for which we presume the
reference frame to be classical, in order to measure the spin
component $\mathbf{S}$ of a particle along a direction $\hat{n}$
we use the projections
\begin{equation}
    P_{n} = \frac{I_2}{2}\pm\hat{n}\cdot\mathbf{S}.
\end{equation}
Now the question arises: what do we mean by a classical reference
frame and in which aspects it is different from a quantum
mechanical reference frame? A QRF is different from its classical
counterpart in two ways. First, due to the inherent uncertainty in
its direction, the measurement results are only an approximation
of what would be obtained using the classical reference frame.
Second, each time the quantum reference frame is used, it suffers
a back-action which causes the future measurements to be less
accurate.

We model the QRF as a spin-$l$ particle, the spin components
described in the normal manner by an operator $\mathbf{L}$, and
consider it being used to make measurements of the direction of a
series of spin-1/2 particles, each described by an operator
$\mathbf{S}$. A measurement of the relative orientation between
the QRF and one particle is given by a measurement of
$J^2=(\mathbf{L}+\mathbf{S})^2$ (the optimal measurement
\cite{BRST} for determining the relative orientation), i.e.
projection onto the $j=l\pm\frac{1}{2}$ irreps as described by projectors
\begin{equation}
    \Pi_{\pm} =\frac{1}{2} \left (I_{2d}\pm\frac{4\mathbf{L}\cdot\mathbf{S}+I_{2d}}{d} \right ),
\end{equation}
with
\begin{equation}
\Pi_{+}+\Pi_{-}=I_{2d}.
\end{equation}
where $d=2l+1$. To verify this works as an approximate measurement
of the particle's spin we then calculate the partial trace over
the reference, initially in a state $\rho$, which yields POVM
operations corresponding to the two outcomes given by
\begin{equation}
\Lambda^\pm_\rho=\Tr_{R}[\Pi_\pm(\rho \otimes I_2)]=
\frac{1}{2}(I_{2}\pm \frac{4 \<\mathbf{L}\>
\cdot\mathbf{S}+I_2}{d}).
\end{equation}
Note that the induced measurement on the source only depends on
the expectation values of angular momentum of the reference frame,
and we can write
\begin{eqnarray}\label{lambda}
\Lambda^+_\rho&=& \frac{l+1}{d}I_2+\hat{n}_\rho\cdot\mathbf{S} \nonumber \\
\Lambda^-_\rho&=&\frac{l}{d}I_2-\hat{n}_\rho \cdot\mathbf{S},
\end{eqnarray}
where
\begin{equation}
\hat{n}_\rho=\frac{\< \mathbf{L}\>}{l+\frac{1}{2}}.
\end{equation}
As is clear, this induced measurement is an approximation of what
we have in (1) such that  as $l$ approaches infinity this
approximation becomes more and more accurate.

After the reference frame has been used to measure a source
particle, it experiences a back-action that can be described as a
quantum channel, or a completely positive trace preserving (CPTP)
map \cite{NC}, which depends on the polarization direction of the
source particles $S$. Note that for the moment we
presume the specific measurement result obtained is ignored. To derive this
map we consider
\begin{equation}
\E[\rho]= \Tr_s[\Pi_+(\rho\otimes\xi)\Pi_+
+\Pi_-(\rho\otimes\xi)\Pi_-],
\end{equation}
in which $\rho$ is the state of the reference frame and $\xi$ is
the state of the source particle. Using the expressions for
$\Pi_{\pm}$, we may express this channel as
\begin{eqnarray}
\E[\rho]&=&(\frac{1}{2}+\frac{1}{2d^{2}})\rho \nonumber\\
 &&+\frac{8}{d^2}\Tr_s[\mathbf{L}\cdot\mathbf{S}
(\rho\otimes\xi)\mathbf{L}\cdot\mathbf{S}]\nonumber \\
&&+\frac{2}{d^{2}} (\rho (\mathbf{L}\cdot\mathbf{\langle
S\rangle})+\rho(\mathbf{L}\cdot\mathbf{\langle S\rangle}))
\end{eqnarray}
This expression is coordinate independent and as such
 we can choose to introduce a background frame in which the source particles have their spin aligned along the $Z$-axis. In this case the state of the sources is given by $\xi=\frac{1}{2}(I+z \sigma_z)$ so that $\<S_z\> =z/2$ and $\<S_x\> = \<S_y\> =0$, and
\begin{eqnarray}
\E[\rho]&=&(\frac{1}{2}+\frac{1}{2d^{2}})\rho+\frac{2}{d^{2}}\sum_{i=x,y,z}L_{i}\rho
L_{i} \\ \nonumber &+& \frac{z}{d^2}(L_z\rho+\rho L_z+L_+\rho
L_--L_-\rho L_+)
\end{eqnarray}
This can be written in the more illuminating form:
\begin{eqnarray}\label{Emap1}
\E[\rho]&=&(\frac{1}{2}+\frac{1-z^2}{2d^2})\rho \nonumber \\
&&+\frac{2}{d^2}(L_z+z/2)\rho(L_z+z/2) \nonumber \\
 &&+ \frac{1+z}{d^2}L_+\rho L_- +\frac{1-z}{d^2}L_- \rho L_+.
\end{eqnarray}
As shown in \cite{PY}, the reference frame to leading order
suffers a drift in its orientation due to non-zero polarization in
the measured particles. This drift tends to align the reference
frame with that of the stream of polarized source particles and
constitutes an equilibrium condition in the absence of
depolarization effects.

To analyse the relative orientation between the QRF and the source
particles we consider an orthonormal frame $(\hat{x}',\hat{y}',
\hat{z}')$, obtained from the Cartesian frame
$(\hat{x},\hat{y},\hat{z})$ via a rotation, which transforms
$(L_x, L_y, L_z)\rightarrow
(L_x'(\mathbf{\theta}),L_y'(\mathbf{\theta}),L_z'(\mathbf{\theta}))
$ such that $\<L_x'(\mathbf{\theta}) \> =\<L_y'(\mathbf{\theta})
\>=0$ and $\<L_z'(\mathbf{\theta}) \>=rl$ for some fractional $r$.
Here $r$ quantifies the polarization of the quantum reference
frame, which is aligned along the direction $\hat{z}'$. Since, by
symmetry, the QRF will remain in the $X$-$Z$ plane, the
transformation is a rotation about the $Y$-axis and takes the form
$L_x'(\theta) = L_x \cos \theta -L_z \sin \theta,L_y'(\theta) =
L_y$ and $L_z'(\theta) = L_z \cos \theta +L_x \sin \theta$. In
\cite{PY} it was shown  that in the limit of large $l$ the map
(\ref{Emap1}) can be approximated to $\O(1/l)$ as
\begin{equation}\label{Erot}
\E [\rho]\approx \rho+i\frac{rz}{2l}\sin\theta [L_y,\rho],
\end{equation}
where $\theta$ is the angle between the polarization of the
sources ($Z$-axis) and the polarization of the reference frame.
Consequently, the measurement process produces an average rotation of the
reference frame through an angle $\Omega (\theta) = -\frac{rz}{2l}
\sin \theta$ towards the polarization direction of the sources.

\subsection{Beyond the Average Map}
Equation (\ref{Emap1}) provides the evolution of the reference
frame due to a measurement process in which we discard the actual
measurement outcome, and represents the average evolution of the
reference frame. However we obtain a more accurate evolution if we
take into account the specific sequence of measurement outcomes.

The average map $\E[\rho]$ can be written as
\begin{equation}
\E[\rho]=p_{+}\E_+[\rho]+p_-\E_-[\rho].
\end{equation}
where a $\pm$ outcome occurs with probability $p_\pm(\theta)$ and
the QRF evolves according to
\begin{eqnarray}
\E_\pm[\rho]&=& \Tr _s [ \Pi_\pm (\rho \otimes \xi ) \Pi_\pm ]
/p_\pm
\end{eqnarray}
or more explicitly,
\begin{eqnarray}
&&p_\pm \E_\pm[\rho] = \left (\frac{1}{4}\pm
\frac{1}{2d}+\frac{1-z^2}{4d^2} \right )\rho
\pm \frac{z}{2d}(\rho L_z+L_z \rho) \nonumber\\
&&
\hspace{-0.3cm}+\frac{1}{d^2}(L_z+\frac{z}{2})\rho(L_z+\frac{z}{2})
+ \frac{1+z}{2d^2}L_+ \rho L_- + \frac{1-z}{2d^2}L_- \rho
L_+\nonumber
\end{eqnarray}
These maps may be approximated to $\O(1/l)$ as
\begin{eqnarray}
\E_\pm[\rho]&\approx& \frac{\rho}{2p_{\pm}} \pm\frac{z}{4lp_{\pm}}
(\rho L_{z}+L_{z}\rho)
 + i\frac{zr}{4lp_{\pm}}\sin{\theta} [L_y,\rho]\nonumber
\end{eqnarray}
where the probability of a plus or minus outcome is $p_\pm(\theta)
= \frac{1}{2}\pm\frac{1}{2}zr\cos \theta$ for $l \gg 1$.

Recall that we have defined the angle of inclination of the QRF in
terms of vanishing expectation values, in particular the relation
$\Tr[L'_{x}(\theta_\pm)\E_\pm[\rho] ] = 0$ will define the angle
$\theta_\pm$ that the transformed state $\E_\pm [\rho]$ makes with
the $Z$-axis. While on the other hand, $\Tr[L'_{x}(\theta)\rho ] =
0$ defines the initial angle $\theta$. Since $\Omega_\pm =
\theta_\pm - \theta$ we find that $\Omega_\pm$ is determined from
the relation
\begin{eqnarray}
&&\frac{\sin\Omega_{\pm}}{\sin(\Omega_\pm +\theta)}+\frac{zr}{2l}\cos \Omega_\pm \nonumber\\
&&\pm\frac{z}{2rl^2}\left [2\<
L^2_z\>-\cot(\Omega_{\pm}+\theta)\<\{ L_z,L_x\}\> \right ]=0
\end{eqnarray}

The unusual terms are the quadratic expectation values in the
square brackets, which indicate that the dynamics depends on
reference frame observables beyond simply the polarization. After
many measurements the dependence on these observables will tend to
cancel on average, however for a small number of measurements
their influence is of importance.

The polarizations of the source particle and the QRF together
define a distinguished frame, which is described by the triple
$(L'_x(\theta), L'_y(\theta), L'_z(\theta) )$. In this natural
frame we find that

\begin{eqnarray}\label{o-opm-rel}
\tan \Omega_\pm \hspace{-0.1cm}&=&\hspace{-0.1cm}
\frac{-\frac{zr}{2l}\sin\theta \pm\frac{z}{rl^2}(\cos\theta
\<L'_x(\theta)L'_z(\theta)\>-\sin\theta
\<L'_x(\theta)^2\>)}{1+\frac{zr}{2l}\cos\theta\pm\frac{z}{rl^2}(\cos\theta
\<L'_z(\theta)^2\>-\sin\theta
\<L_x'(\theta)L_z'(\theta)\>)},\nonumber
\end{eqnarray}
where we
have used that the transformed angular momentum operators obey the
usual $su(2)$ commutation relations, $[L'_i (\theta), L'_j
(\theta) ] = i\epsilon_{ijk} L'_k(\theta)$ \cite{Rose,Edmonds}. We
now consider two interesting classes of states, for which more
explicit analytic solutions for $\Omega_\pm$ exist.

\subsubsection{Partially Coherent States}
Since a distinguished frame exists for which the QRF is initially
in a state for which $\<L'_x(\theta)\> =\<L'_y(\theta)\> = 0$ and
$\<L'_z(\theta)\>=rl$, we can restrict to a class of states with
the property that the initial state $\rho$ obeys
\begin{eqnarray}\label{assumption}
\Tr[ \rho L'_i (\theta) L'_j (\theta) ] = \<l, rl| L_i L_j |l,rl
\>
\end{eqnarray}
for any choice of $i$ and $j$. These states possess a high degree
of symmetry about their axis of polarization and include, as a
special case, coherent states.  On this set of states we obtain
$\Omega_\pm$ in a form that only depends on its initial angle of
inclination $\theta$,
\begin{eqnarray}\label{opm}
\Omega_{\pm}&=&-\arctan \left [\frac{z\sin\theta(r^2 \pm [l(1-r^2)
+1])}{2rl(1 \pm zr \cos \theta)}  \right ].
\end{eqnarray}

For $r=\pm 1$ we have perfectly coherent states and find that
$\Omega_\pm$ vanishes in the $l \rightarrow \infty$ limit, as
expected, and the QRF becomes a fixed classical reference frame.
Indeed for this perfectly coherent state we find that $\Omega_-=0$
for all theta, which occurs since the rank of the corresponding
projector is $2l+2$ and the initial state lies entirely in its
support.

 However for $-1<r<1$ we see that as $l \rightarrow \infty$ the rotation angles $\Omega_\pm$ are non-zero, in contrast to the average map. We find that
\begin{eqnarray}
\lim_{l\rightarrow \infty} \Omega_{\pm}&=&\pm\arctan \left [
\frac{z(r^2-1)\sin\theta}{2r(1\pm zr \cos \theta)} \right ]
\end{eqnarray}
which reflects that the QRF does not have perfect polarization
along its axis.

Indeed, from (\ref{lambda}) it can be seen that for $\<
\mathbf{L}\> \cdot \mathbf{S}= rl S'_z(\theta)$ in the limit $l
\rightarrow \infty $ the source particles do not undergo a perfect
projective measurement, but instead are subject to a `fuzzy
measurement' with POVM operators $\Lambda_\rho^\pm= (1/2)(I \pm 2r
\hat{n}\cdot \mathbf{S})$.

For the QRF, the large transverse fluctuations in
$\<L'_x(\theta)^2\>$ are affected by the projection $\Pi_{l \pm
1/2}$ and leads to a non-vanishing asymptotic rotation of the QRF.

\begin{figure}[t]
\centering
\includegraphics*[width=11.0cm, bb=3cm 8cm 22cm 21cm]{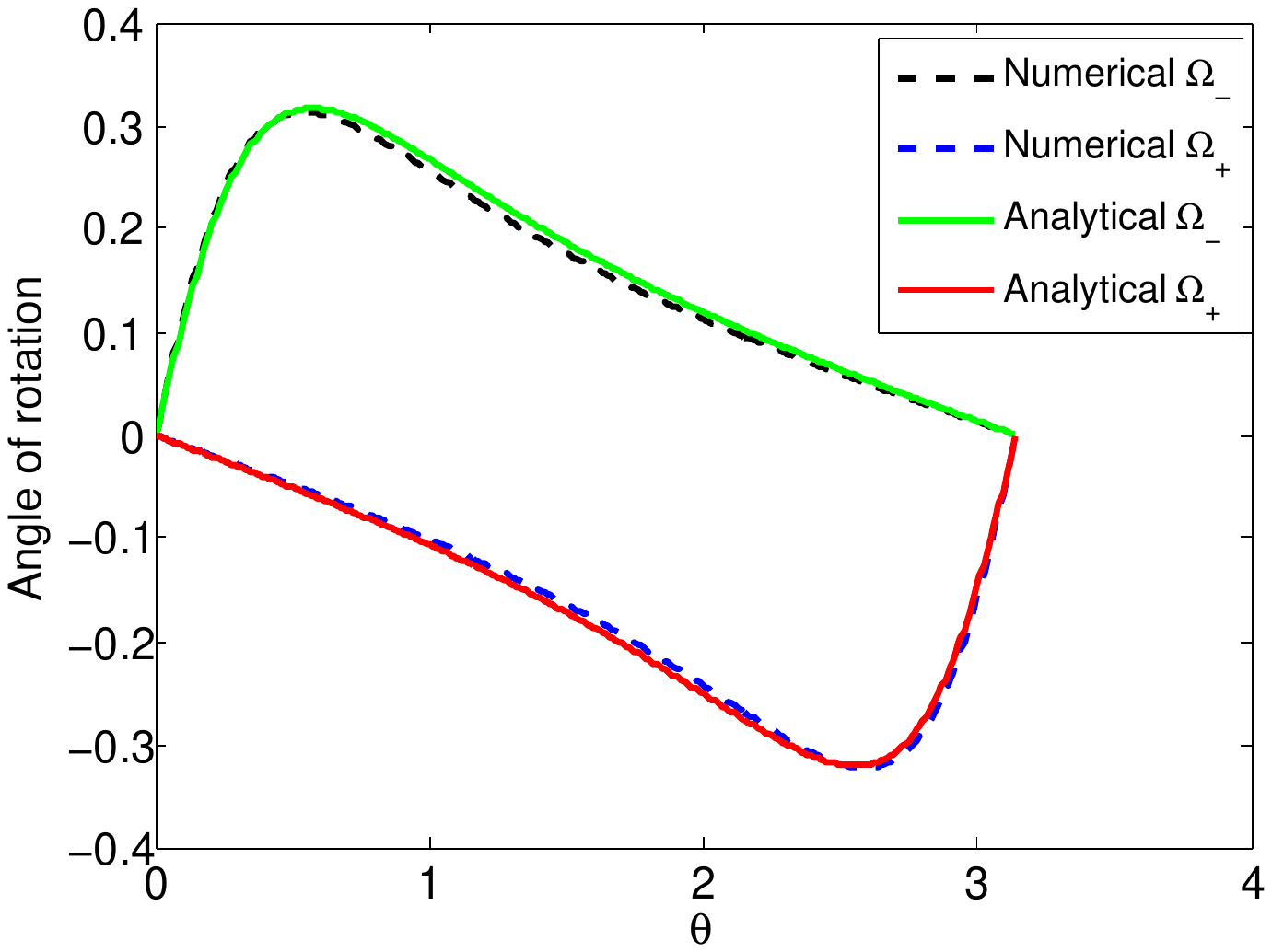}
\caption{(color online) A comparison between a numerical simulation of the
rotation produced by the map $\E_\pm$ on a family of mixed states
of the form (\ref{mixedinitial}) with the expression $\Omega_\pm$
obtained in equation (\ref{opm}). The state that we have
considered in this figure is $\rho=p \exp [ - i \beta L_y]
|l,10\>\< l,10| \exp [ i \beta L_y]+ (1-p)\exp [ - i \beta L_y]
|l,40\>\< l,40| \exp [  i \beta L_y]$ with $p=0.2$ and $L=100$.}
\label{Opmcompare}
\end{figure}

 In figure (\ref{Opmcompare}) we compare our analytical expression (\ref{opm}) with numerical results for  a set of mixed initial states of the form
\begin{eqnarray}\label{mixedinitial}
&& \hspace{-0.8cm} \rho = p \exp [ - i \beta L_y] |l,k_1\>\< l,k_1| \exp [ i \beta L_y] \nonumber \\
&&+ (1-p)\exp [ - i \beta L_y] |l,k_2\>\< l,k_2| \exp [  i \beta
L_y],
\end{eqnarray}
and find excellent agreement. Indeed this analytic
expression provides a reasonably robust approximation, allowing
for a few percent mixing of a random state to the pure partially
coherent states. In such cases the analytic expressions tend to slightly
overestimate the angles of rotation.

 A convenient subset of these partially coherent states are given by  $\rho = \exp [ -\beta(r) L'_z(\theta)] /Z$, where $Z = \Tr [\exp [ -\beta(r) L'_z(\theta)]$. These states correspond to a QRF partially polarized at an angle $\theta$ to the source particles and with $r=-\frac{1}{l} \partial_\beta \log Z$. These states are special in that they are the highest entropy states subject to these two conditions on $\theta$ and $r$.

\subsubsection{Quadratic Bloch States}

In general, for an $N$-dimensional irrep of $su(N)$ the generators
$\{L_i\}$, together with the identity operator, span the space of
quantum states and so any state admits a `Bloch state' form $\rho
= aI +  \sum_i b_i L_i$. These states have similar properties
\cite{ritter} to the standard Bloch states of a qubit.

However, for a spin-$l$ irrep of $su(2)$ these hermitian operators
no longer span the set of quantum states. Instead, we must use
symmetric polynomials in the generators of the $su(2)$ algebra to
span the full set of states. Futhermore, for any spin-$l$ irrep
there exists a minimal order polynomial expansion (e.g. for
$l=1/2$ the minimal order is 1). Consequently, any truncated
expansion to a lower order will only span a subset of the full
space of states.

A potentially interesting set of states for the spin-$l$ irrep of
$su(2)$, are `Quadratic Bloch States' obtained from a quadratic
combination of $su(2)$ generators
\begin{eqnarray}\label{qbs}
\rho  &=& \frac{1}{2l+1} \left (I + \mathbf{R} \cdot \mathbf{L} +
\frac{1}{2}\sum_{a,b} T^{ab}\{L_a,L_b\} \right ).
\end{eqnarray}
The vector $\mathbf{R}$ and the tensor $T^{ab}$ must obey certain
conditions in order that $\rho$ be a positive trace one operator,
in particular, $T^{ab}$ is a real, symmetric, traceless second
rank tensor. Only for $l=1/2$ and $l=1$ does this expansion cover
the whole set of states.

For such quadratic states, we may calculate an explicit form for
$\Omega_\pm$ using certain trace identities. The quadratic terms
$\<L'_i(\theta) L'_j(\theta) \>$ receive non-zero contributions
from the the $T^{ab}$ components only. They are determined
explicitly using the identity
\begin{eqnarray}
\Tr [\{L_i ,L_j \} \{L_k,L_m \} ] &=& \alpha_l \delta_{ij}\delta_{km} \nonumber \\
&&+\beta_l (\delta_{ik}\delta_{jm}+\delta_{im}\delta_{jk})
\end{eqnarray}
where the coefficients $\alpha_l$ and $\beta_l$ are given by
\begin{eqnarray}
\alpha_l &=&\frac{l(l+1)(2l+1)(1+2l(l+1))}{15}  \nonumber \\
\beta_l &=&\frac{l(l+1) (4l^2-1) (2l+3)}{15}.
\end{eqnarray}

Repeated use of this trace identity gives us that the angles of
rotation for this family of states are given by
\begin{eqnarray}\label{qbsangle}
\tan \Omega_\pm \hspace{-0.1cm}&=&\hspace{-0.1cm}
-\frac{15zr^2\sin\theta \pm z(l+1)(d^2-4)T_1(\theta)}{30rl\pm
z(l+1)(d^2-4)T_2(\theta)},
\end{eqnarray}
with $d=2l+1$ and
\begin{eqnarray}
T_1(\theta) &=&T^{xx} \cos \theta \sin 2\theta - T^{zz} \sin \theta \cos 2\theta + T^{xz} \cos 3\theta \nonumber \\
T_2(\theta) &=&T^{xx} \sin \theta \sin 2\theta + T^{zz} \cos
\theta \cos 2\theta + T^{xz} \sin 3\theta\nonumber
\end{eqnarray}
being the contributions from the quadratic order terms in the
state.

As already mentioned, these Quadratic Bloch States are generally a
subset of all quantum states. For $l=1/2, 1$ this expansion covers
the full set of states, however the analytic expressions for the
rotation angles is a poor approximation since we are neglecting
$\O (1/l^2)$ terms. As we increase $l$ the set of states described
by (\ref{qbs}) becomes a smaller and smaller fraction of all
states. In addition the net polarization $r$ of these states is
generally small and this means that the analytic expressions
obtained are still very approximate. It is expected that by
including higher order terms that contribute to the net
polarization $r$, but do not contribute to the quadratic
expectation values, the expression (\ref{qbsangle}) would have
greater accuracy. We leave this issue for a future investigation.

\section{Evolution of the Reference Frame under a Unitary Interaction}

Single spin-qubit rotations are typically performed using an
external classical field that can be considered as some large
amplitude coherent state within the quantum description. In
practice the finiteness of the external control field - equivalent
to our reference system - means that the qubit and the field
become entangled, resulting in a slightly imperfect rotation of
the qubit. This was investigated for the case of a 2-level atom
interacting with a single cavity mode initially in a coherent
state in \cite{kimblevanenk}. Our model is very similar - our
reference spin is essentially starting in a large amplitude
spin-coherent state. We are interested, however, in the case that
it is reused multiple times for applying single qubit rotations to
different qubits. As there is no other reference system it is
clear the interaction hamiltonian should be rotationally
invariant, that is, it should depend only on the relative
orientations of the qubit and the frame. The most natural choice
is to consider a coupling Hamiltonian of the form $\mathbf{L}\cdot\mathbf{S}$,
which, in the limit of large $l$, would yield a standard single
qubit unitary rotation on the spin.

We consider therefore that the QRF and each incoming spin are
coupled for a time $t$ such that the evolution takes the form
$e^{i\mathbf{L}\cdot\mathbf{S}t}$. As already discussed, the
sequential measurement of total angular momentum causes the
reference frame to rotate in the $X$-$Z$ plane, in other words the
expectation value of the y-component of the QRF is always zero
during the whole process, however we shall see the unitary
interaction produces a rotation around an axis that depends on the
precise duration of the interaction.

\subsection{Backreaction on the quantum reference frame}

First we write the unitary $e^{i t \mathbf{L}\cdot\mathbf{S}}$ in
a simpler form. For this purpose we use the equations,
\begin{eqnarray}    J^{2}&=&(l+\frac{1}{2})(l+\frac{3}{2})\Pi_{+}+(l-\frac{1}{2})(l+\frac{1}{2})\Pi_{-}\nonumber\\
   I_{2d} &=& \Pi_+ + \Pi_-
\end{eqnarray}
and obtain that $\mathbf{L}\cdot\mathbf{S}=\frac{1}{2} (l \Pi_+ -
(l+1) \Pi_-)$. It is clear from this expression that, in the $l
\rightarrow \infty$ classical limit,  coherent interactions
with a highly polarized QRF induces rotation about the spatial
axis defined by the observable $Z= \Pi_+ - \Pi_-$, while for
finite $l$ we have that $U = \Pi_+ + e^{-i \gamma} \Pi_-$ where
$\gamma =t(l +1/2)$.

The effect that the QRF suffers due to a single unitary
interaction $U(\gamma)$ is then given by the CP map
 \begin{eqnarray}
\F_\gamma[\rho]&=&\Tr_s [ U(\gamma) (\rho \otimes \xi) U(\gamma)^\dagger ] \nonumber \\
&=&\Tr_s[(\Pi_+(\rho\otimes\xi)\Pi_+]+\Tr_S[(\Pi_-(\rho\otimes\xi)\Pi_-]\nonumber\\
&&+ e^{-i \gamma} \Tr_s [\Pi_-(\rho\otimes\xi)\Pi_+] +e^{i \gamma}
\Tr_s [\Pi_+(\rho\otimes\xi)\Pi_-].\nonumber
 \end{eqnarray}

\begin{figure}
  \includegraphics*[width=11.0cm, bb=2cm 8cm 22cm 23cm]{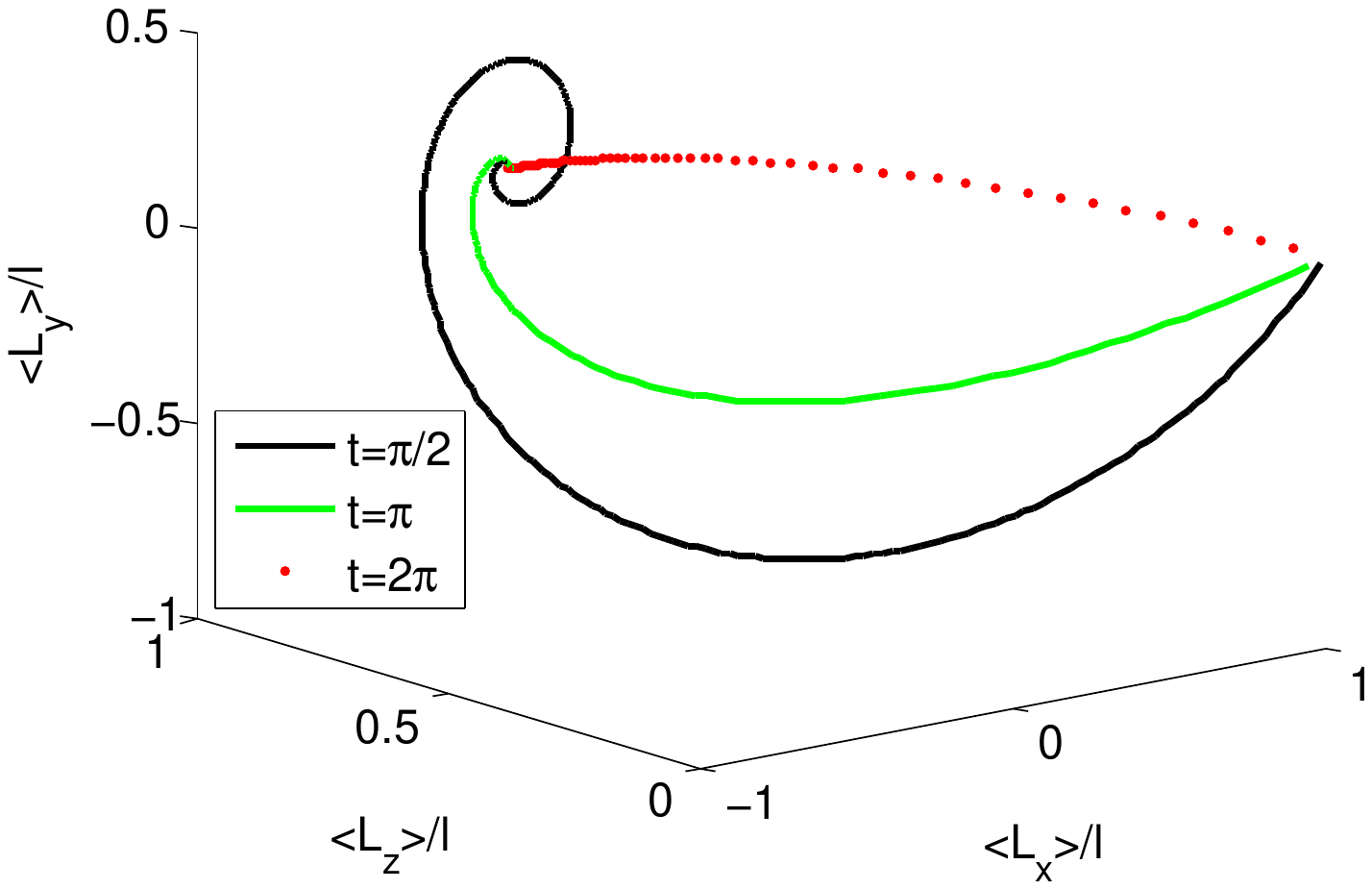}
\caption{$\langle L_{x}\rangle /l$, $\langle L_{y}\rangle /l$ and
$\langle L_{z}\rangle/l$. The rotation induced on the reference
frame due to the unitary interaction with the source particle for
$l=16$. The source particles are polarized along the z-axis with
z=1 and the QRF initially points along the x-axis, $\theta=\pi/2$.
In this figure N=500 source particles has been used.}
\label{unitaryrotation}
\end{figure}

Once again we assume a source particle polarized along the
$Z$-axis and in the state $\xi = \frac{1}{2} (I + z \sigma_z)$ and
obtain that
\begin{eqnarray}
\F_\gamma [ \rho] &=&\frac{1}{2d^2} (d^2+1 +(d^2-1)\cos \gamma)\rho +  \nonumber \\
&&\hspace{-1.5cm} +\frac{4}{d^2} \sin^2 \frac{\gamma}{2}\sum_\alpha L_\alpha \rho L_\alpha +i z \frac{4}{d^2}\sin^2 \frac{\gamma}{2}(L_y\rho L_x - L_x \rho L_y) \nonumber\\
&&\hspace{-1.5cm}+\frac{2z}{d^2}\sin^2 \frac{\gamma}{2}(L_z\rho +
\rho L_z)+i\frac{z}{d}\sin \gamma [L_z,\rho ],
\end{eqnarray}
from which we only keep up to $\O(1/l)$ terms to obtain the
following expression for the effect of the unitary interaction on
the reference frame:
\begin{eqnarray}\label{Frot}
\hspace{-0.4cm} \F_\gamma[\rho]
\hspace{-0.1cm}&\approx&\hspace{-0.1cm} \rho + \frac{izr }{l}\sin
\theta \sin^2 \frac{\gamma}{2} [L_y, \rho] + \frac{iz }{2l}\sin
\gamma [L_z,\rho].
\end{eqnarray}
This induces a linear transformation of the initial polarization
vector $(\<L_x\> , \<L_y\>, \<L_z\>)$ sending it to $(\<L_x\>_\F ,
\<L_y\>_\F, \<L_z\>_\F)$ where $\<L_i\>_\F \equiv \Tr[\F_\gamma
[\rho] L_i]$, and the new components are given by
\begin{eqnarray}
\<L_x\>_\F&=&\<L_x\>+\frac{z}{2l}\sin \gamma \<L_y\> - \frac{rz}{l}\sin \theta \sin^2 \frac{\gamma}{2}\<L_z\>\nonumber\\
\<L_y\>_\F &=&\<L_y\> -\frac{z}{2l}\sin \gamma \<L_x\>  \nonumber\\
\<L_z\>_\F &=& \<L_z\> -\frac{rz}{l}\sin\theta \sin^2
\frac{\gamma}{2}\<L_x\>.
\end{eqnarray}
To order $\O(1/l)$ this is a rotational map around the axis
$(0,\frac{1}{r} \csc \theta \cot \frac{\gamma}{2}, 1)$ through an
angle $\Omega_\F (\gamma, \theta) = \frac{z}{l} \sin
\frac{\gamma}{2} \sqrt{ r^2 \sin^2 \theta \sin^2 \frac{\gamma}{2}+
 \cos^2 \frac{\gamma}{2} }$, and in particular it is clear that $\lim_{l \rightarrow \infty} \Omega_\F (\gamma, \theta) =0$. This rotational dynamics is illustrated in Fig.~\ref{unitaryrotation}, where we perform repeated coherent interactions between the QRF and a stream of source particles.

\section{Correcting the Drift of a Quantum Reference Frame}

In this section we consider certain approaches that allow us to
correct the drift of the reference frame due to the projective
measurement $\{ \Pi_\pm \}$.

If, in addition to the source of particles $S$, which are aligned
in the $Z$-direction, we also have access to another set of
particles $\bar{S}$, which are aligned in the $-Z$-direction, then
our intuition is that we may recover the quadratic scaling of
\cite{BRST} by alternating the measurements on systems from $S$
with measurements on systems from $\bar{S}$. Since the sequence of
measured particles has zero net polarization no directional drift
of the QRF occurs.

However, this approach requires the use of an equal number of
`corrective' $\bar{S}$ particles as measured particles - but is
this the optimal strategy to eliminate drift? Two different
strategies present themselves, but before discussing them we first
establish an operational criterion for the usefulness of the QRF.

\subsubsection{Operational Criterion}
We wish to define an operational criterion by which to judge how
well the finite-sized QRF does in the task of mimicking a
projective measurement on the source particles.

To judge the quality of the measurement we follow \cite{BRST} and
consider the probability of successfully finding the correct
result $l+\frac{1}{2}$ when the test particle is pointing along
$+\hat{n}$ (the initial direction of the reference frame)or
finding the correct result $l-\frac{1}{2}$ when the test particle
is pointing along $-\hat{n}$:
\begin{eqnarray}
P_{\mbox{\tiny succ}} &=& \frac{1}{2}
\Tr[\Pi_+(\rho\otimes|\hat{n}\>\<\hat{n}|)+\Pi_{-}(\rho\otimes|-\hat{n}\>\<-\hat{n}|)]\nonumber
\\
&=& \frac{1}{2}(1+\hat{n}\cdot\hat{n}_\rho).
\end{eqnarray}

\begin{figure}
  \includegraphics*[width=11.0cm, bb=3cm 8cm 22cm 21cm]{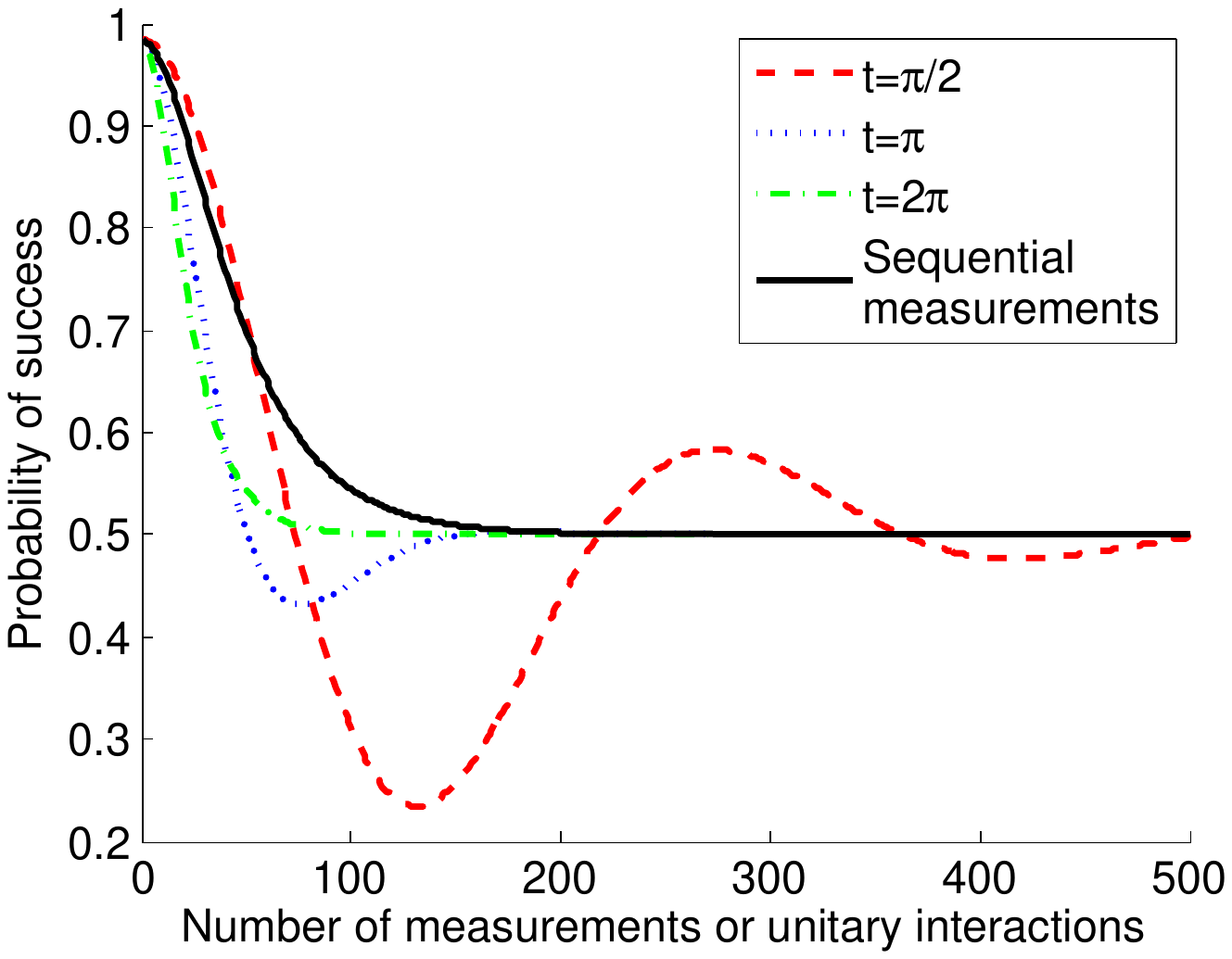}
\caption{$P_{\mbox{\tiny succ}}$ as a function of the number of
interactions for the case in which source particles are polarized
along the z-axis ($z=1$) and the QRF is initially in the coherent
state $l=16$ pointing along the x-axis, i.e.
$\theta=\pi/2$.}\label{meas-int}
\end{figure}

In \cite{BRST} it was shown that the number of measurements a QRF
could be used for before $P_{succ}$ falls below some threshold
scaled quadratically with $l$ if the source of particles was
unpolarized. In \cite{PY} it was shown that the scaling becomes
only linear with $l$ if the source of particles being measured has
some net polarization. In Fig.~\ref{meas-int} we show the
degradation of the reference frame under a sequence of either
measurement interactions (solid line) or unitary interations for various
values for $\gamma$.

\subsection{Correction via Unitary Interactions}

 The first corrective mechanism we consider is to make two measurements of particles from $S$ and then to implement a unitary $U=e^{-i2\pi\mathbf{L}\cdot\mathbf{S}}$ between the QRF and a particle from $\bar{S}$.

\begin{figure}
  \includegraphics*[width=11.0cm, bb=3cm 8cm 22cm 21cm]{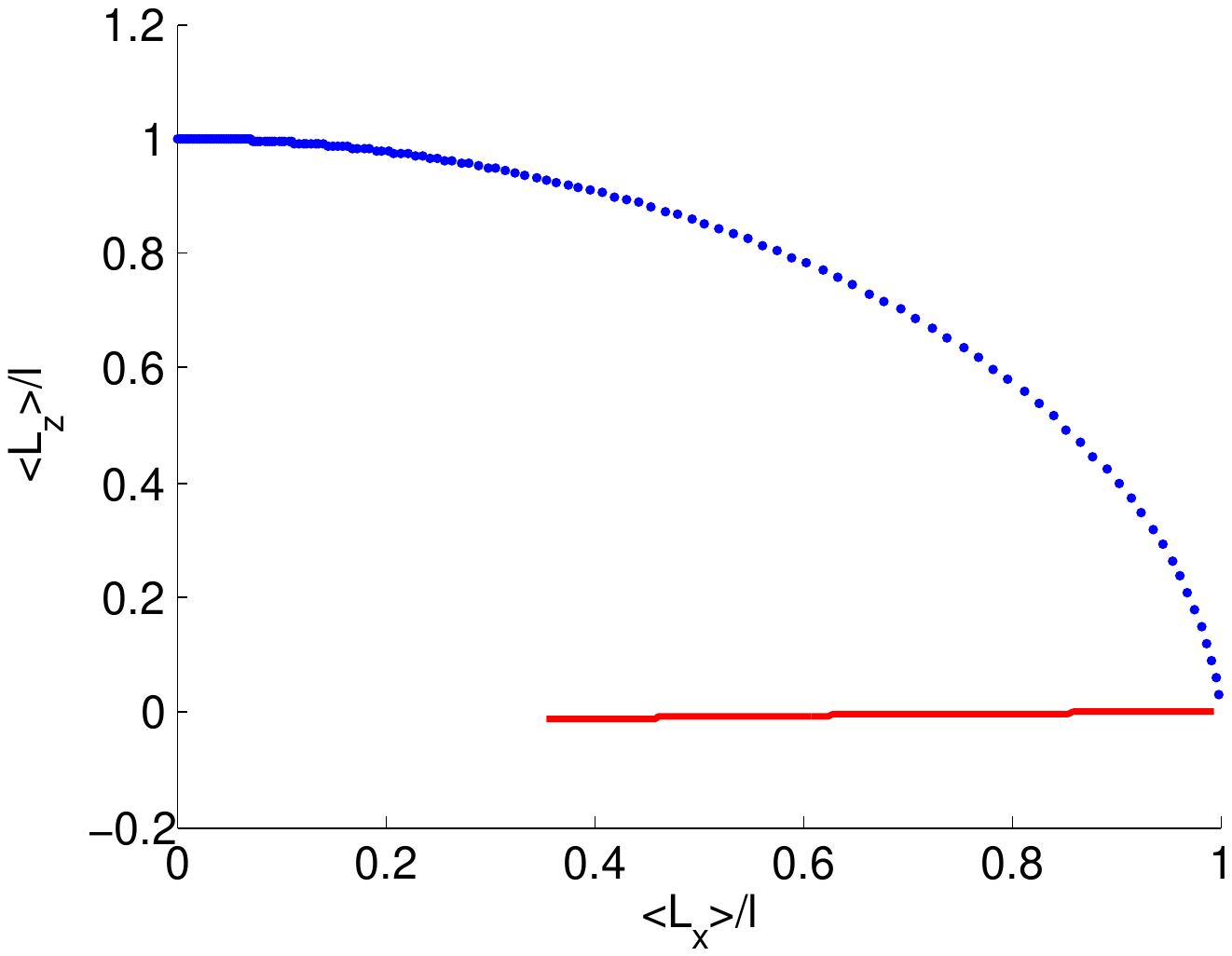}
\caption{$\< L_{z}\>/l$ vs. $\< L_{x}\> / l$ for $l=16$. The
source particles are polarized along the z-axis and the QRF is
initially in the coherent state pointing along the x-axis. The
blue dotted line corresponds to the case of sequential
measurements and the red dotted line is for the case of unitary
interaction $e^{i2\pi \mathbf{L}\cdot\mathbf{S}}$ after two
measurements.} \label{Lz vs Lx}
\end{figure}

In Fig.~\ref{Lz vs Lx} we plot the $Z$-component of angular
momentum of the QRF versus its x-component. The blue line is the
degradation with no correction, as considered in \cite{PY}. The
red line is for the case in which we have applied the unitary
mentioned above after every \emph{two} measurements - we observe that
this method helps us to essentially completely correct the
rotation of QRF (the drift towards the polarization of $S$).

To understand why this works, we see from equation (\ref{Frot}) that the unitary interaction can
generate a rotation about the $Y$-axis of $\frac{rz}{l} \sin
\theta \sin^2 \gamma/2 $. For the particular choice of
$\gamma=\pi$ we have that the unitary interaction produces a
rotation exactly twice as large as the measurement interaction,
while maintaining the reference frame in the $X$-$Z$ plane. By
using a source particle from $\bar{S}$ we can ensure that this
rotation acts in the opposite direction to the drift to
equilibrium, and it is easily checked that
\begin{eqnarray}
\F_\pi[\E^2[\rho]]=\rho+\O(1/l^2).
\end{eqnarray}
An important point to emphasize is that the application of the
unitary interaction not only can correct the polarization drift to
$\O(1/l^2)$, but it does so without requiring knowledge of the
relative angle $\theta$ between the QRF and the source particles.

For very large $l$ we have greater freedom regarding when in the course
of a sequence of $N$ measurements the corrective unitaries are
performed. If $1 \ll N \ll l$, then we have that $p_\pm(\theta)$
is roughly constant over the course of $N$ measurements. The
actual measurement sequence is highly probable to be a typical
measurement sequence with $p_+N$ plus outcomes and $p_-N$ minus
outcomes. However, since $N \ll l$ the QRF has rotated through a
total angle $p_+N\Omega_+ +p_-N\Omega_- = N\Omega$, which may be
corrected with $N/2$ unitary interactions distributed arbitrarily
between the $N$ measurements.

\begin{figure}
  \includegraphics*[width=11.0cm, bb=3cm 8cm 22cm 21cm]{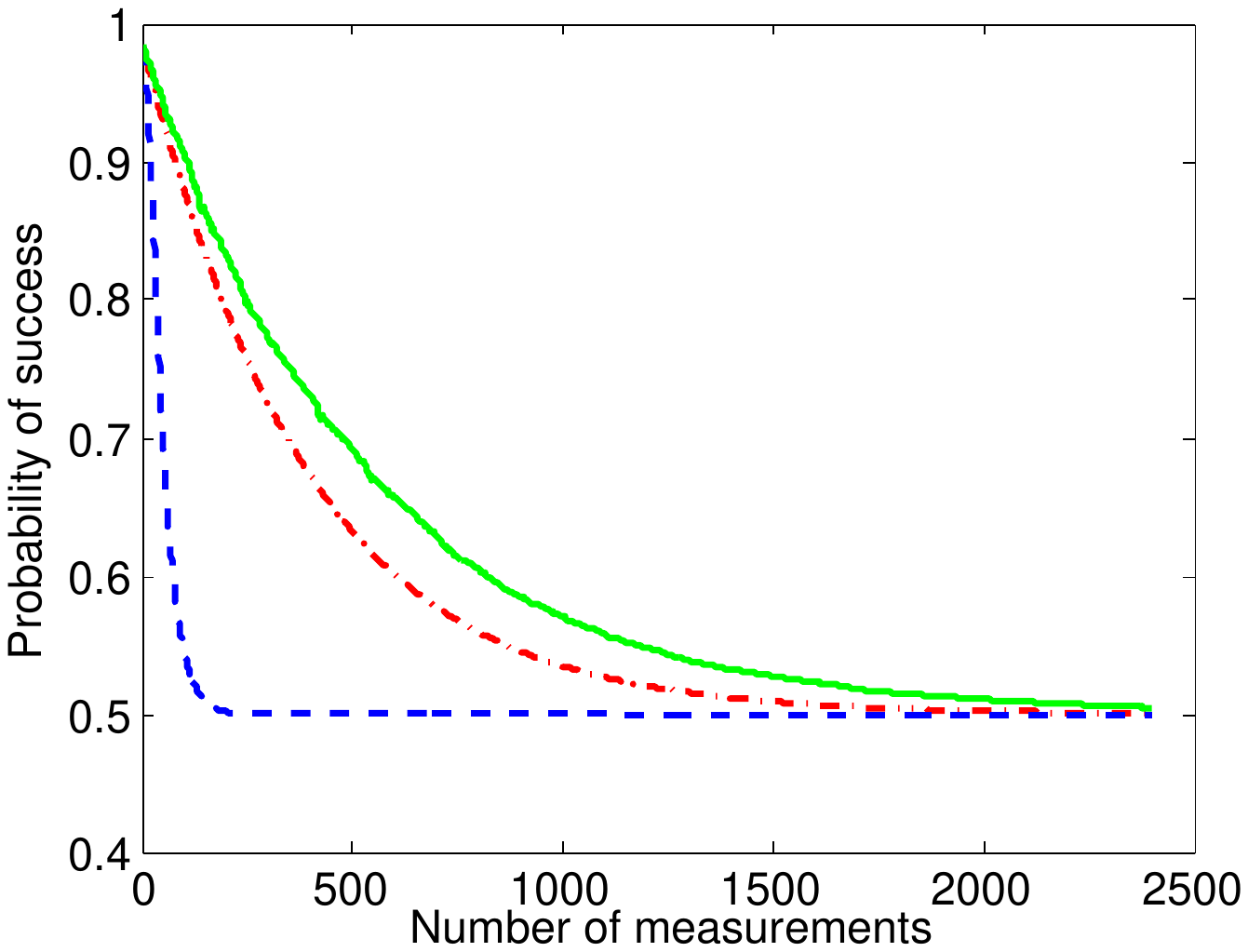}
\caption{A comparison of probability of success for obtaining
correct measurement result in three different cases for $l=16$.
The dashed line corresponds to the case of sequential
measurements, the dashed-dotted line is for the case in which we
correct the measurement result via applying unitary interactions
after two measurements and the solid line belongs to the case of
correction via applying unitary interactions after each plus
outcome.}\label{Probsucc}
\end{figure}

In Fig.~\ref{Probsucc}, $P_{succ}$ is plotted against the number
of measurements for the two cases mentioned above. We can clearly
see that the longevity of the QRF is now improved. In this figure
the horizontal axis is for the number of measurements and the
particles used to improve the probability of success are not
included, so with the use of particles from $\bar{S}$ we may
extend the lifetime of the QRF to $\O(1/l^2)$ in a more efficient
manner than described in the previous section.



\subsection{Keeping track of measurement results}

None of the work on QRF degradation has considered the option of
keeping track of the measurement results. This has been primarily
for the sake of maintaining a simple pedagogy. We can now consider
the possibility of actively feeding back individual measurement
results to correct the frame's drift.

With probability $p_+$ the QRF is transformed as $\rho \rightarrow
\E_+ [\rho]$ and similarly with probability $p_-$ the QRF is
transformed as $\rho \rightarrow \E_- [\rho]$. A measurement
history for the reference frame may be described via
$\vec{s}=(s_1,s_2, \dots, s_N)$, with $s_i=\pm$. This sequence of
outcomes in term corresponds to a evolution of the QRF given by
$\E_{\vec{s}} [\rho] := \E_{s_N} [ \cdots \E_{s_2}[
\E_{s_1}[\rho]]]$.

The probabilities for large $l$ are given by $p_\pm (\theta) =
\frac{1}{2} (1 \pm zr \cos \theta )$ where $z$ is the polarization
of the source particles and $l$ is the polarization of the of the
reference frame, as described earlier. Since we are considering
$\O(1/l)$ effects we shall assume that $r$ is approximately
constant for $N\ll l$.

 Note that in the context of the above measurement history, the probabilities for each outcome $s_i$ are not independent since $p_\pm$ has angular dependence and so depends on previous rotations induced by $s_{i-1}, s_{i-2}, \dots$.

We may again use the unitary interaction, however unlike the case
of the average map $\E$, no simple correction exists for an
individual plus or minus outcome for two reasons. Firstly, the
angle of rotation generated by the unitary interaction decreases
monotonically with $l$ and so fluctuations, such as the ones
discussed earlier, may be much too large to correct.

Secondly, the unitary rotation goes sinusoidally with the relative
angle $\theta$ between the source particles and the QRF, while the
rotations due to the individual outcomes are in general
complicated functions of $\theta$. A knowledge of $\theta$ would
be needed to tune the unitary interaction correctly. However, it
should be that any auxilary background reference frame that we may
introduce should not feature in the experimental considerations,
and should serve only as a useful intermediate construct.
\textit{`Information is physical'}, and so any meanful coordinate
system must be associated with an actual physical system.

Of course, one could take the view that an large background system
already exists, and relative to this we have already determined
the angles of inclinations of both the source particles and
quantum reference frame, and hence know the value for $\theta$.
However, in this case, the goal of considering unitary corrections
would then be to preserve the known state of the QRF in between
measurements, as distinct from providing a reliable reference
frame with which one determines the unknown relative angle with an
ensemble of source particles through repeated measurements.

With a knowledge of the relative angle $\theta$ we may tune the
unitary interaction appropriately, using either a source particle
from $S$ or $\bar{S}$, and correct sufficiently small rotations of
the QRF. However, in the event of large measurement rotations, the
best we can do between individual measurements would be to perform
the largest allowable rotation in the required direction -
numerics indicate that for the two projective outcomes $\Pi_\pm$
we can always correct one outcome entirely and the other for
$\pi/2 < \theta < \pi$.

\section{Discussion and Outlook}

In this paper we have analysed in some detail the induced dynamics
of a quantum reference frame as it is used to measure the spins of
a sequence of source particles, and also used to implement unitary
interactions on the source particles.

We found that the average behaviour of the QRF is to gradually
rotate into alignment with the source particles at an $\O (1/l)$
rate. If we pay attention to the induced dynamics subsequent to a
particular measurement outcome, we find that the dynamics is not
so simple and large fluctuations can exist, which depend on
observables quadratic in $\mathbf{L}$. We considered the
restriction to a simple class of initial states for which the
dynamics depends purely on the inclination of the QRF relative to
the source particles. For such states we found that fluctuations
may persist even in the infinite limit, and which give non-trivial
dynamics. Of course in this limit there is, on average, no net
rotation of the QRF.

We found that by performing a unitary interaction between the QRF
and source particles every third step, we could eliminate the
$\O(1/l)$ directional of the reference frame under the average
map.

Future work might include the issue of parameter-estimation on the
state of the source particles. While ordinary projective
measurements possess a degeneracy between the polarization of the
source particles and the relative angle between the QRF and the
particles, the presence of dynamics breaks this degeneracy and
potentially allows a richer measurement inference.

In the ideal projective measurement case, the measurement
probabilities are given by $p_\pm = (1/2)(1 \pm z \cos \theta)$,
and so doing a sequence of measurements only gives us the value of
$z \cos \theta$. However, in the presence of dynamics, the
reference frame responds differently to the polarization $z$ of
the source and to the relative angle $\theta$ with the source. For
example, by allowing the QRF to gradually come into alignment with
the source particles the measurement pattern is eventually
determined solely by $z$, while the early-time outcomes encode the
dependence on $\theta$. Such a separation of parameters is a
result of the non-trivial dynamics of the finite quantum reference
frame.

It is also possible to do parameter estimation plus correction in
parallel. Initially we know nothing of $\theta$ and so can take it
to lie uniformly between $0$ and $\pi$.  However, for example,
getting a string of many plus outcomes implies that the relative
angle $\theta$ is quite small. Each successive measurue outcome we
obtain allows us to update our estimate for $\theta$ and in each
case we can use our best estimate to perform a unitary correction,
ideally converging in on a stable distribution and the correct
value for the relative angle.

Alternatively, in the event that we are ignorant of the relative
angle $\theta$ it may be possible to perform a `conditional'
corrective unitary interaction. The idea is that the source
particle that has been measured with the QRF encodes the relative
angle between the QRF and the unmeasured particles in its new
state. It may be possible to transfer this $\theta$ dependence to
in a manner which improves the corrective
procedures.

Finally, it would be of interest to extend the analysis we have
conducted here to study how measurement and unitary interactions
behave between a large QRF and higher spin particles.


\end{document}